\documentclass{osa-article}

\journal{oe}

\articletype{Research Article}

\definecolor{col1}{rgb}{0.3,0.7,0}

\definecolor{col2}{rgb}{0,0.1,0.7}

\definecolor{col3}{rgb}{0.6,0.2,0.3}

\definecolor{col4}{rgb}{1, 0, 0}

\usepackage{multirow}
\usepackage{braket}

\begin{document}

\title{400\%/W second harmonic conversion efficiency in 14~\textmu m-diameter gallium phosphide-on-oxide resonators}

\author{Alan D. Logan,\authormark{1,*} Michael Gould,\authormark{2}, Emma R. Schmidgall,\authormark{2} Karine Hestroffer,\authormark{3} Zin Lin,\authormark{4} Weiliang Jin,\authormark{5} Arka Majumdar,\authormark{1,2} Fariba Hatami,\authormark{3} Alejandro W. Rodriguez,\authormark{5}   and Kai-Mei C. Fu\authormark{1,2}}

\address{\authormark{1}Department of Electrical and Computing Engineering, University of Washington, Seattle WA 98195\\
\authormark{2} Department of Physics, University of Washington, Seattle WA 98195\\
\authormark{3}Department of Physics, Humboldt-Universitat zu Berlin, 12489 Berlin, Germany \\
\authormark{4} John A. Paulson School of Engineering and Applied Sciences, Harvard University, Cambridge, MA 02138 \\
\authormark{5} Department of Electrical Engineering, Princeton University, Princeton, NJ 08544, USA}
\email{\authormark{*}adlogan@uw.edu}

\begin{abstract*}
Second harmonic conversion from 1550~nm to 775~nm with an efficiency of 400\% W$^{-1}$ is demonstrated in a gallium phosphide (GaP) on oxide integrated photonic platform. The platform consists of doubly-resonant, phase-matched ring resonators with quality factors $Q \sim 10^4$, low mode volumes $V \sim 30 (\lambda/n)^3$,
and high nonlinear mode overlaps. Measurements and simulations indicate that conversion efficiencies can be increased by a factor of 20 by improving the waveguide-cavity coupling to achieve critical coupling in current devices.
\end{abstract*}

\section{Introduction}

Parametric nonlinear processes such as frequency conversion play a critical role in several established and emerging technological applications~\cite{ref:fejer1994nofc,ref:boyd2008nlo}, including ultrashort-pulse shaping~\cite{DeLong94,Arbore97}, generation of light in strategically important spectral windows~\cite{Kuo06,Vodopyanov06,Krischek10}, spectroscopy~\cite{Heinz82,Ozawa08}, and quantum information~\cite{ref:kwiat1995nhi,Vaziri02,Tanzilli05,Zaske12,ref:chang2014qno}.
In particular, efficient frequency conversion of single photons from optically accessible qubits is expected to enable long-range fiber transmission~\cite{ref:pelc2012dcq} and increase compatibility between dissimilar quantum memories~\cite{ref:degreve2012qds}.
Two important practical considerations dictating the performance of quantum networks are scalability and efficiency.  Similar considerations arise in the context of frequency conversion, with typical devices exploiting resonators designed to exhibit high field strengths and long-lived modes to increase efficiency and reduce power requirements~\cite{BravoAbad07,rodriguez2007chi}.
These conditions have been commonly met in macroscopic optical cavities (such as large-etalon resonators~\cite{Furst10,wang2014integrated,lin2017nonlinear,wolf2017cascaded}) and waveguides~\cite{wang2017metasurface,chen2018modal} that sacrifice spatial confinement for increased ability to engineer modes at the desired wavelengths and satisfy the requisite phase matching conditions~\cite{ref:boyd2008nlo}.
However, there has been increased interest in exploring designs that exploit the small footprint, low mode volume, and wider bandwidth offered by integrated photonic devices, which could drastically improve scalability in size, cost, and power consumption~\cite{kuo2014second,buckley2014second,ref:lake2016efficient,roland2016phase,Guo2016,ref:hendrickson2014inp,mohamed2017efficient,rao2018second,moille2018nonlinearities}.
These include whispering gallery mode resonators~\cite{lin2017nonlinear,kuo2014second}, singly-resonant photonic crystal cavities~\cite{li2015design}, and nanoplasmonic and dielectric metasurfaces~\cite{butet2015optical,fryett2017phase,sitawarin2018inverse}.
More recently, integrated ring resonators in aluminum nitride (AlN) on sapphire demonstrated second harmonic generation (SHG) efficiencies as high as 17,000\% W$\mathrm{^{-1}}$~\cite{ref:bruch20181sh}, far exceeding the performance of prior devices.

In this work, we study SHG in a gallium phosphide (GaP) on oxide platform that asphalts the path for achieving high-efficiency frequency conversion in robust, compact, and wide bandwidth integrated cavities.
In particular, GaP-on-oxide ring resonators were designed for quasi-phase-matched SHG from the telecommunication C band (1550 nm) to the near-infrared (775 nm). This work represents at least an order-of-magnitude improvement over prior SHG GaP devices~\cite{ref:lake2016efficient,Rivoire2011} while also extending device functionality and applicability via coupling to on-chip waveguides.
Waveguide-to-waveguide SHG efficiencies of up to 400\% W$\mathrm{^{-1}}$ are observed in 14~\textmu m diameter rings. The conversion efficiency is primarily limited by overcoupling to both input and output waveguides.

In addition to its significant $\mathrm{\chi^{(2)}}$ second-order nonlinear susceptibility ($\sim$100 pm/V~\cite{Corso1996}), gallium phosphide offers a unique combination of properties useful for frequency conversion applications.

GaP has a high index of refraction (n~=~3.31 at 637 nm~\cite{Bond1965}) compared to AlN (n~=~2.2~\cite{Pastrnak}) or most traditional nonlinear materials like $\mathrm{LiNbO_3}$ (n~=~2.28~\cite{Zelmon1997}), allowing fabrication of ultra-low mode volume resonators on a wide variety of substrates, including diamond (n~=~2.4)~\cite{Barclay2009,Schneider2018,Englund2010,Rivoire2011}.
With a wide bandgap $E_\mathrm{G}$ of 2.32 eV, GaP maintains transparency into the visible spectrum, permitting efficient frequency conversion of a wider range of wavelengths than other high-index materials such as GaAs ($E_\mathrm{g}$=1.42 eV, $\mathrm{\chi^{(2)}}\sim 220$~pm/V~\cite{ref:shoji1997ass}). Notably, the emission wavelength of the diamond nitrogen-vacancy center, a solid-state qubit candidate, falls within the transparency window of gallium phosphide~\cite{Englund2010}.

\section{Model and Design}
\label{sec:modeldesign}

At low input powers, the effect of down conversion on the efficiency and power requirements for achieving SHG in a doubly resonant cavity is negligible~\cite{rodriguez2007chi}. In this undepleted regime, the conversion efficiency corresponding to a cavity supporting modes at angular frequencies $\omega_{1,2}$ and coupled to waveguides, is given by~\cite{rodriguez2007chi,bi2012high}:
\begin{equation}
\frac{P_{2,out}}{P_{1,in}^2} = \frac{|\chi ^{(2)}|^2}{\epsilon _0 \lambda_1^3} | \bar{\beta} |^2 \frac{2}{\omega _1} \frac{Q_1^4 Q_2^2}{Q_{c1}^2 Q_{c2}} = \frac{|\chi ^{(2)}|^2}{\epsilon _{0} \lambda_1^3} | \bar{\beta} |^2 \frac{2}{\omega _1} Q_{i1}^2 Q_{i2} \frac{Q_{i1}^2 Q_{c1}^2}{(Q_{c1}+Q_{i1})^4} \frac{Q_{i2} Q_{c2}}{(Q_{c2}+Q_{i2})^2},
\label{eq:conveff}
\end{equation}
Here, $Q_{k}$, $Q_{ik}$, and $Q_{ck}$ denote the loaded, intrinsic, and coupling quality factors of mode $k=\{1,2\}$, where mode 1 is the fundamental and mode 2 is the second harmonic. The coefficient $\mathrm{\bar{\beta}}$ is the dimensionless nonlinear overlap of the fundamental and second harmonic modes,
\begin{equation} \label{eq:beta}
\bar{\beta} = \frac{\int_\mathrm{NL} \sum_{i \neq j \neq k} (E_{1i} E_{2j}^{*} E_{1k} + E_{1i} E_{1j} E_{2k}^{*})  d\boldsymbol{r}}{\left( \int \epsilon_1 |E_1|^2 d\boldsymbol{r} \right) \sqrt{\int \epsilon_2 |E_2|^2 d\boldsymbol{r}}} \sqrt{\lambda_1^3},
\end{equation}
a generalization of the familiar phase matching figure of merit~\cite{ref:boyd2008nlo}. For zincblende crystals such as GaP, $\mathrm{\chi^{(2)}}$ is nonzero only for mutually perpendicular field components.~\cite{ref:boyd2008nlo}. Note that the integral in the numerator is only evaluated over the extent of the nonlinear medium. Notably, achieving high conversion efficiencies requires large $\mathrm{\chi^{(2)}}$, high intrinsic quality factors, critical coupling at both the fundamental and second-harmonic modes, and high nonlinear overlap $\mathrm{\bar{\beta}}$. Previous work with GaP photonics on diamond~\cite{ref:gould2016lsg} has indicated that the intrinsic quality factor is determined by process-dependent sidewall roughness, so the device design process was focused on achieving high mode overlap and coupling rather than increasing $Q$.

A ring resonator topology was chosen to allow control over mode properties and independent coupling to on-chip waveguides with minimal design variables. To ensure sufficient confinement at 1550~nm, a 427~nm thick photonics layer of (100) GaP on a thermal $\mathrm{SiO_2}$ substrate was used. The ring structure imposes rotational symmetry on resonator modes. However, in (100) GaP, the effective nonlinear susceptibility changes sign every $\mathrm{90^{\circ}}$, so quasi-phase-matching is required to avoid back-conversion. Also, because only mutually perpendicular field components contribute to $\mathrm{\bar{\beta}}$ in GaP, the fundamental mode must be transverse-electric (TE) and the second harmonic mode must be transverse-magnetic (TM). To allow straightforward device optimization under these constraints, Eq.~\ref{eq:beta} was translated to cylindrical coordinates:
\begin{align} \label{eq:beta_ring}
\bar{\beta} &= \sqrt[]{\lambda_1^3} \int_{0}^{2\pi} \beta^{+} e^{i(2m_1 - m_2 + 2)\theta} + \beta^{-} e^{i(2m_1 - m_2 - 2)\theta} d\theta\  \\
\begin{split}
\beta^{\pm} &= \frac{1}{( \int \epsilon_1 |E_1|^2 ~ r ~ drdz ) ~ \sqrt[]{\int \epsilon_2 |E_2|^2 ~ r ~  drdz}} \int_{NL} \bigl (2[E_{1r} E_{1z} (E_{2r}^{*} + E_{2\theta}^{*}) + E_{1\theta} (E_{1r}  E_{2z}^{*} + E_{1z} E_{2r}^{*})] \\
&\pm i[(E_{1r}^2 - E_{1\theta}^2) E_{2z}^{*} + E_{1z} (E_{1r}  E_{2r}^{*} - E_{1\theta} E_{2\theta}^{*})]\bigr) \ r \ drdz \nonumber
\end{split}
\end{align}
where $m_k$ is the azimuthal mode number of mode $k$. In this coordinate system, it is readily aparent that $\mathrm{\bar{\beta}}$ vanishes if the quasi-phase-matching condition $2m_1 = m_2 \pm 2$ is not satisfied. Due to the relatively small $\mathrm{\theta}$ field component, both $\beta^{+}$ and $\beta^{-}$ depend primarily on $E_{1r}^2 E_{2z}^{*}$, with the main contributions to $\beta^{+(-)}$ weighted toward the outside (inside) edges of the ring.

The resonator design process was based on two-dimensional eigenmode simulations of a ring cross-section. First, the fundamental $\mathrm{TE_{00}}$ at $\mathrm{\lambda_1 = 1550~nm}$ was simulated for a ring of radius on the order of ten microns. Higher-order TM modes with similar effective index were simulated for $\lambda_2 = 775$~nm, and the mode with the highest $\mathrm{| \beta^{+/-}} |$ was selected. The ring width and radius was then modified via a gradient algorithm to minimize the phase-mismatch parameter $(|2m_1-m_2|-2)$ for the selected modes. Through this process, high-efficiency resonator design candidates could be found with minimal computational resources and without bias toward higher-order harmonic modes. A final ring design of width $w = 840$~nm and radius $r = 7.14$~\textmu m (measured from the center of the waveguide) was found, corresponding to an antisymmetric $\mathrm{TM_{03}}$ second harmonic mode shown in Fig.~\ref{fig:transmission} (inset), which yields $2m_1 - m_2 =  +2$ and $\mathrm{\bar{\beta} = 1.43 \times 10^{-4}}$.

To allow testing of large device arrays on a single chip, the ring resonator was evanescently coupled to waveguides terminating in nearby grating couplers, as shown in Fig.~\ref{fig:layout}a. Critical coupling of both modes is vital for maximizing conversion efficiency, so independent wraparound coupling regions were designed for each ring mode via supermode analysis. Waveguide widths, ring-to-waveguide separations, and coupling region lengths were co-optimized to theoretically provide coupling quality factors of $Q_{1c} \approx \mathrm{2 \times 10^5}$ for the 1550~nm mode and $Q_{2c} \approx \mathrm{2 \times 10^4}$ for the 775~nm mode.

\section{Fabrication and Testing}

\begin{figure}[h]

\centering\includegraphics[width=13cm]{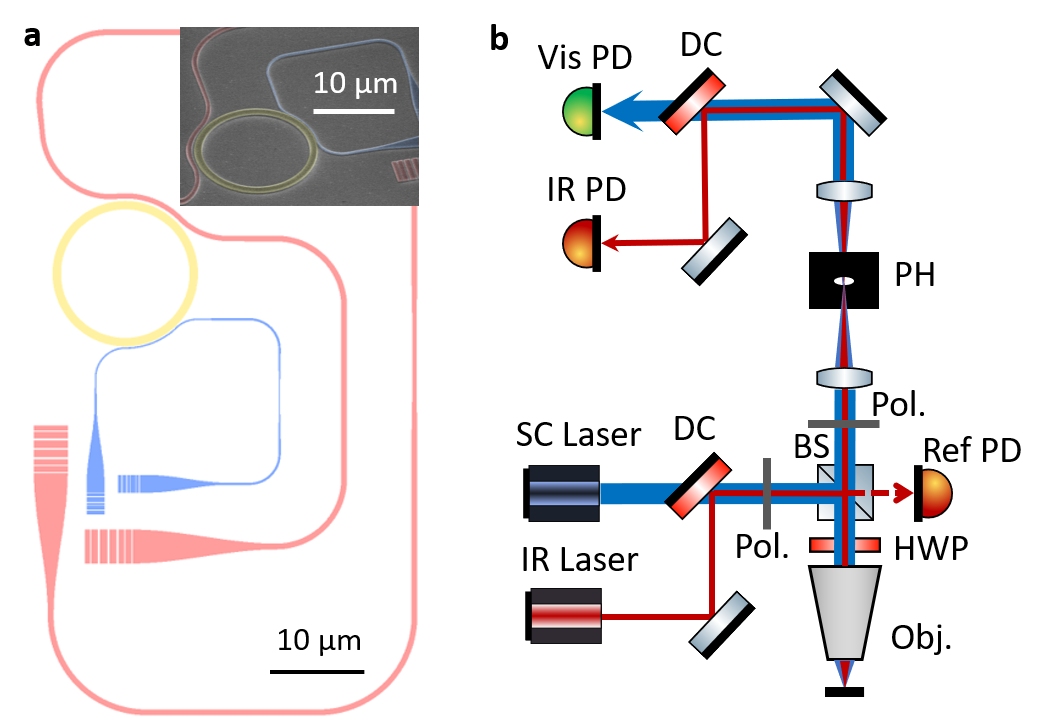}
\caption{(a) On-chip layout of the nonlinear ring resonator (yellow) coupled to two independent input/output waveguides for 775~nm (blue) and 1550~nm (pink) light. The proximity of the grating couplers allow any combination of inputs and outputs to be focused or collected simultaneously by a single microscope objective. {\it Inset:} SEM image of a fabricated GaP SHG device. (b) Free-space measurement setup for the device.  Cross-polarization and the pinhole (PH) are used to eliminate reflected input light. PD: photodiode, Obj: objective, BS: beamsplitter, DC: dichroic mirror, HWP: half-wave plate } \label{fig:layout}
\end{figure}

A 427~nm-thick GaP layer on a 300~nm Al$_{0.8}$Ga$_{0.2}$P sacrificial layer was grown by molecular beam epitaxy on a GaP substrate. A (2.5 mm)$^2$ area GaP membrane was released from the substrate and transferred to a 10 \textmu m thermal SiO$_2$-on-Si substrate. Before transfer, the oxide surface was cleaned and treated with hexamethyldisilazane vapor. The GaP membrane was released from the sacrificial AlGaP layer in 3:100 HF:H$_2$O and transferred to DI water. The membrane was then captured on a water droplet on the oxide substrate. A drying step at 80~$^\circ$C completed the membrane transfer. The described transfer process was used due to its compatibility with transfer to mm-scale diamond chips for quantum information applications using the process described in Refs.~\cite{ref:thomas2014wis,ref:gould2016eez,ref:schmidgall2018fcs}. Recently, wafer-scale GaP membrane transfer to silicon oxide has been realized by other groups via direct wafer bonding followed by substrate removal~\cite{Schneider2018,ref:wilson2018igp}.  

In our devices, the resulting GaP-on-oxide chip was patterned with electron beam lithography, using $\sim$100-nm-thick HSQ as a resist. A final Cl/Ar/$\mathrm{N_2}$ (1.0/6.0/3.0 sccm) reactive ion etch step was used to transfer the mask into the GaP-on-oxide substrate. ~\cite{ref:gould2016lsg}. A device schematic and SEM image are shown in Fig.~\ref{fig:layout}a. Two grating-coupled input/output waveguides, one for the fundamental and one for the second harmonic, are used to couple to the device. A 50-device array was fabricated varying the ring waveguide width from 839~nm to 847~nm to ensure quasi-phase matching and doubly resonant enhancement can be attained even in the presence of fabrication tolerances. The measured ring waveguide widths of the two devices (SHG01, SHG02) exhibiting doubly resonantly enhanced SHG are listed in Table 1.

\begin{table}[]\caption{SHG device characteristics. {\it w} is the resonantor waveguide width. $T$ is the transmission on-resonance.
Uncertainty in $w$ denotes the range of measured values. $Q$ and $T$ are determined by a Lorentzian fit with uncertainty representing the 95\% confidence interval.}
\centering\begin{tabular}
{|cccccc|}
\hline
 Device & $w$  & $Q_1$ & $Q_2$ & $T_1$ & $T_2$\\ \hline
 SHG01 & $847\pm23$~nm & $26,500\pm1500$ & $13,600\pm5400$ & $0.44\pm0.10$ & $0.50\pm0.05$  \\ \hline
 SHG02 & $847\pm17$~nm & $40,700\pm10700$ & $16,800\pm3200$ & $0.81\pm0.05$ & $0.52\pm0.02$ \\ \hline
\end{tabular}
\label{table:table}
\end{table}

\begin{figure}[h]
\centering\includegraphics[width = 0.8\textwidth]{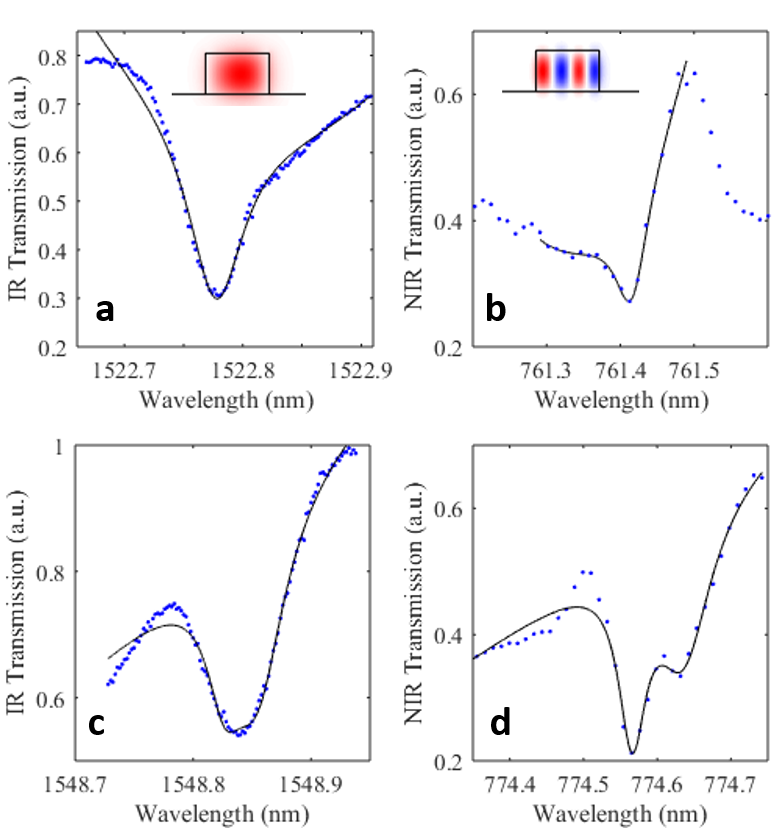} 
\caption{The transmission dip from telecom $\mathrm{TE_{00}}$ (a,c) and near-infrared $\mathrm{TM_{03}}$ (b,d) resonances in devices SHG01 (a,b) and SHG02 (c,d), along with fitted Lorentzian curves. A cross-section of the mode profile is inset. The telecom resonance was measured on an infrared power meter with a tunable laser input, and the near-infrared resonance was measured using a supercontinuum laser and spectrometer.}\label{fig:transmission}
\end{figure}

The devices were tested using the setup shown in Fig.~\ref{fig:layout}(b). A scanning 1550 nm laser (Santec TSL-510) was used to excite the fundamental mode. Fig.~\ref{fig:transmission}(a) shows the telecom transmission measurement curve for SHG01. The Lorentzian fit to the the resonance dip corresponds to $Q_1= 2.65\times10^4$, with $Q_1$ denoting the total quality factor, $1/Q_1 = 1/Q_{c1} + 1/Q_{i1}$. The transmission coefficient on resonance is $T_1 = 0.44$. Transmission spectra of a broadband 775~nm source (supercontinuum laser or LED), shown in Fig.~\ref{fig:transmission}(b), were used to determine the second-harmonic mode quality factor $Q_2$ and transmission $T_2$. Quality factors and transmission coefficients for both modes and both devices are given in Table~\ref{table:table}.  Additionally, a cross-sectional mode profile for both modes are included as insets in Fig.~\ref{fig:transmission}a,b.

Single wavelength transmission measurements at 1550 nm (775 nm) were used to measure the grating coupler efficiency as well as bulk transmission through the microscope set-up at each wavelength. A description of the grating coupler and microscope loss characteristics can be found in Appendix~\ref{appendix:calibration}. The telecom (SHG) grating couplers were designed for 33\% (24\%) efficiency at 1550 nm (775 nm). The measured efficiencies at the experimental resonances were 22\% at 1523~nm (3.2\% at 761~nm) in SHG01 and 19\% at 1549~nm (7.5\% at 774~nm) in SHG02.  All measured values are derived by assuming identical efficiencies for the input and output gratings. 

For SHG measurements, the telecom input grating is used to excite the fundamental mode. Input power is continually monitored by a reference photodiode. Simultaneously, the SHG signal is collected from one of the SHG grating ports (both are tested). Any detuning between the fundamental and SHG excitation wavelengths and their respective resonant modes reduces conversion efficiency. To realize mutual resonance, the device is tested on a temperature-controlled stage. Heating the device causes both resonances to redshift at different rates, allowing relative tuning. SHG conversion efficiency is measured as a function of input wavelength at multiple temperatures to find the maximum conversion efficiency.

\section{Results and Discussion}
\label{sec:Results}

\begin{figure}[h]
\centering\includegraphics[width=\textwidth]{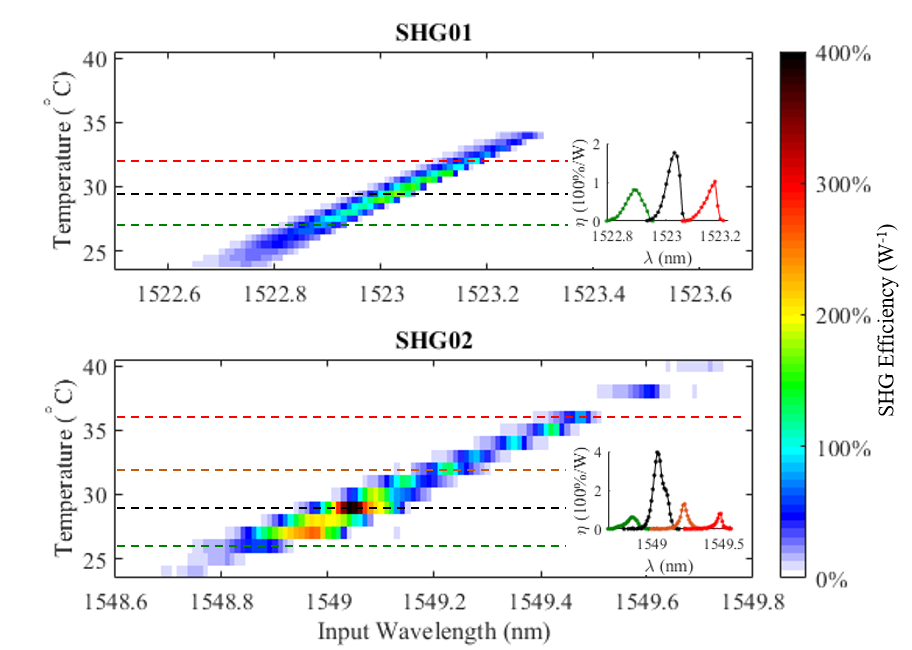}
\caption{(a) SHG conversion efficiency of device SHG01 as a function of both temperature and input wavelength. Conversion efficiency profiles at 27 (green), 29.5 (black), and $\mathrm{32~^{\circ}C}$ (red) are inset. (b) Conversion efficiency of SHG02, with profiles at 26 (green), 29 (black), 32 (orange), and $\mathrm{36~^{\circ}C}$ (red). Asymmetry from thermal bistability is visible in the conversion efficiency profiles of both devices and becomes more pronounced with stage temperature and input laser power. Due to resonance splittings, SHG02 exhibits additional asymmetry as well as efficiency peaks at multiple temperatures.}
\label{fig:Temp_Sweep_Overview}
\end{figure}

\begin{center}
\begin{figure}[h]
\includegraphics[width=\columnwidth]{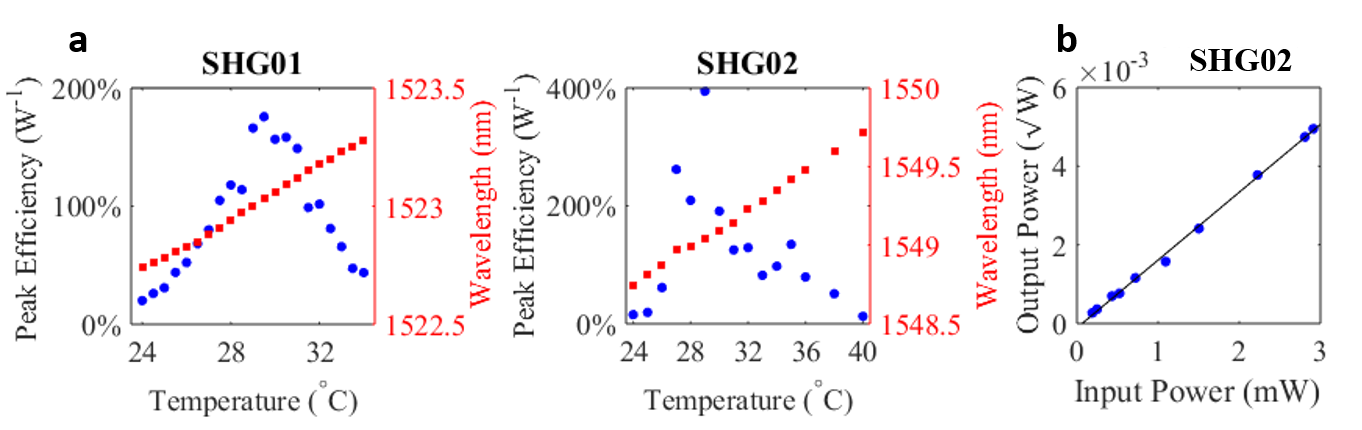}
\caption{(a) Maximum SHG efficiency as a function of temperature for both devices. The red squares are the corresponding fundamental wavelength as a function of temperature. (b) Square root of the SHG output power as a function of fundamental input power, showing the expected linear dependence. Both input and output powers are calculated in-waveguide powers.}
\label{fig:Temp_Sweep_Pwr}
\end{figure}
\end{center}

Quasi-phase matched resonances were found in device SHG01 at a fundamental wavelength of 1523.1~nm and in SHG02 at 1549.1~nm. The quasi-phase matching condition was identified by the strong and highly temperature-dependent second harmonic conversion of light at the fundamental resonance, as shown in Fig. \ref{fig:Temp_Sweep_Overview}. In devices in which only single-resonance enhancement is observed, both the conversion efficiency and the effect of temperature are far weaker.
A maximum waveguide-to-waveguide conversion efficiency $\eta$ of 175\% $\mathrm{W^{-1}}$ was observed in SHG01. $\eta$ includes SHG signals propagating in both directions of the SHG waveguide. The maximum conversion efficiency as a function of temperature followed a Lorentzian profile with a full-width-at-half-maximum of $\sim 4~^\circ$C. (Fig.\ref{fig:Temp_Sweep_Pwr}a), with the peak efficiency wavelength redshifting linearly to follow the fundamental resonance (Fig.~\ref{fig:Temp_Sweep_Pwr}a). The assymmetrical shape of the SHG efficiency curves (inset Fig.~\ref{fig:Temp_Sweep_Overview}a) is attributed to a redshift of the resonance as it is heated by the laser.

Device SHG02 exhibited both a higher maximum efficiency of 400\% $\mathrm{W^{-1}}$ as well as a more complex dependence on the temperature and fundamental wavelength (Fig.~\ref{fig:Temp_Sweep_Overview}b). 
Due to a splitting of both the fundamental and second harmonic resonances~\cite{ref:little1997sri}, SHG02 exhibited efficiency peaks at multiple temperatures instead of the single peak seen in SHG01. The double-humped structure in the peak efficiency curve (black curve, inset of Fig.~\ref{fig:Temp_Sweep_Overview}b) is attributed to these split resonances. 	

Fig.~\ref{fig:Temp_Sweep_Pwr}(c) shows the SHG efficiency, measured from only one SHG grating, as a function of the fundamental input power on double resonance for SHG02. As expected, the SHG power increases quadratically with the fundamental power, with no sign of depletion at waveguide input powers of up to 3~mW.  The peak waveguide-to-waveguide conversion efficiency is 0.84\% in device SHG02. Increasing the waveguide input power beyond 3~mW resulted in thermal optical bistability~\cite{ref:almeida2004obs}, causing a discrete hop of the resonance when tuning the fundamental on resonance. This relatively low absolute conversion efficiency points to the practical need to increase the per input power efficiency for quantum conversion applications as well as a limit on the total SHG power that may be produced for classical applications.

To compare the experimental device performance to the simulations (Eq.~\ref{eq:conveff}), it is necessary to know the coupling $Q_c$ and intrinsic $Q_i$ of both modes. Based on transmission measurements, the fundamental and second harmonic resonances of both devices are not critically coupled, a condition that is met when $Q_c = Q_i$, and $T = 0$. The measured finite $T$ gives us only the magnitude of the difference between $Q_c$ and $Q_i$, $T = |Q_i - Q_c|^2/(Q_i + Q_c)^2$. Rewriting Eq.~\ref{eq:conveff} in terms of the $T_1$ and $T_2$, we obtain
\begin{equation}
	\eta_{\mathrm{theory}} = \frac{P_{2,out}}{P_{1,in}^2} = \frac{|\chi ^{(2)}|^2}{\epsilon _0 \lambda_1^3} | \bar{\beta} |^2 \frac{2}{\omega _1} Q_1^2 Q_2 \left(\frac{1 \pm \sqrt[]{T_1}}{2}\right)^2 \frac{1 \pm \sqrt[]{T_2}}{2},
\end{equation}
in which the $+$ signs correspond to the case where both modes are overcoupled, $Q_c > Q_i$, and the $-$ signs to the case where both modes are undercoupled $Q_c < Q_i$. Using the measured $Q$'s and $T$'s in Table~\ref{table:table} and the calculated $\mathrm{\bar{\beta}}$ from Sec.~\ref{sec:modeldesign}, we are only able to obtain reasonable agreement between experiment and theory in the case where both modes are overcoupled.
The SHG efficiency in this case is $\eta_{\mathrm{theory, oc}}^{\mathrm{SHG01}} = 230\%~\mathrm{W}^{-1}$ and $\eta_{\mathrm{theory, oc}}^{\mathrm{SHG02}} = 425\%~\mathrm{W}^{-1}$ and corresponds to the highest theoretical efficiency. For SHG02, $\bar\beta$ (Eq.~\ref{eq:beta}) is calculated for a standing wave instead of a traveling wave due to the observed resonance splitting.

To further investigate the coupling regime, the coupling regions of both devices were imaged and measured by scanning electron microscopy and simulated by finite-difference-time-domain and supermode analysis (Appendix~\ref{appendix:coupling}). Within the measurement uncertainty, the coupling quality factor for the 775~nm mode could be as low as $1.2 \times 10^3$ and as high as $1 \times 10^6$, so overcoupling of this mode is plausible. The lowest reasonably attainable coupling $Q_{1c}$ for the 1550~nm mode was $1.5 \times 10^5$, significantly higher than the measured loaded quality factors. This analysis indicates the fundamental mode is most likely undercoupled, in which case theoretical calculations predict much lower conversion efficiencies than observed. Due to this uncertainty in the coupling factors, we are unable to reconcile the experimentally measured efficiencies with these simulations at this time. However, theoretical calculations do not include the effect of surface roughness or sidewall angles so other coupling mechanisms may be a factor.

\section{Conclusion and Outlook}

In summary, we observe near 400\%/W SHG conversion efficiency in waveguide-integrated GaP resonators. The high conversion efficiency is achieved with resonant enhancement of both the fundamental and second harmonic modes, meeting the quasi-phase matching requirement, and achieving high mode-overlap in small mode-volume structures. These experimental results indicate two areas in which device performance can be immediately and significantly improved. First, the coupling should be decreased to achieve critical coupling. Assuming intrinsic quality factors $Q_i$ of $10^5$ (consistent with our current measurements), the expected SHG efficiency exceeds 8000\%/W. Reducing the diameter of these simple rings to 5 \textmu m should increase $|\bar\beta|^2$ by a factor of 3. Under such a regime of operation, pump depletion would occur at modest powers $<< 1$~mW, avoiding heating-induced optical bi-stability. Critical coupling to such small rings may prove challenging, but recent theoretical results utilizing inverse design methods have demonstrated single waveguide couplers capable of achieving critical coupling at multiple frequencies~\cite{ref:jin2018idc}. Finally, the ability to achieve higher nonlinear coupling factors $\bar\beta$ in ring resonators is largely hampered by their diminishing capacity to confine light with decreasing sizes. While photonic crystals and associated structures can overcome such a tradeoff~\cite{BravoAbad07}, they typically can only do so over narrow bandwidths.  Recently proposed inverse design strategies~\cite{Lin2016,sitawarin2018inverse} point a way toward new kinds of multi-mode cavities capable of confining light at disparate wavelengths in ultra-small volumes, and exhibiting  orders of magnitude larger $\bar\beta$ factors, the subject of ongoing experimental efforts.

\appendix
\setcounter{figure}{0}
\section{Measurement Calibration}
\label{appendix:calibration}

All reported SHG conversion efficiencies are on-chip efficiencies, i.e. based on input and output power in the on-chip waveguides. On-chip powers were derived from off-chip efficiency measurements.
Grating coupler efficiency was measured by comparing the transmission spectrum from each coupling photonic circuit to a reflection spectrum from the thermal oxide substrate. In Fig.~\ref{fig:layout}a, the telecom coupling circuit is pink and the SHG/near-infrared coupling circuit is blue. Near-infrared spectra were measured using a supercontinuum laser for excitation and detected by a grating spectrometer with a CCD detector. Telecom spectra were measured using a scanning laser for excitation and a power meter for detection.  

Excitation polarization was adjusted using a  $\mathrm{\lambda/2}$ plate directly before the objective to excite TE (1550 nm) or TM (775 nm) modes. Reflected excitation light was filtered from the collection path using both a cross-polarizer and a spatial filter (a pinhole) to select the output grating coupler. The efficiency of a single grating was calculated from the power transmission spectrum assuming identical input/output grating couplers and negligible on-chip losses. This assumption of identical couplers is reasonable for SHG01 in which the SHG efficiency is similar when exciting either telecom grating coupler. There is a temperature-dependent discrepancy for SHG02 which is attributed to the splitting of the resonances due to backscattering~\cite{ref:little1997sri}. Near the resonances of interest, the efficiency of the telecom grating coupler for device SHG01 (SHG02) was measured as 22\% (19\%). Efficiency of the near-infrared grating was 3.2\% (7.5\%). Transmission measurements through all off-chip optical components were conducted (depicted in Fig.~\ref{fig:layout}b.) The total telecom power delivered to grating coupler input is $0.31\times P_{\mathrm{RefPD}}$ in which $P_{\mathrm{RefPD}}$ is the power measured on the reference photodiode. Including the grating coupling efficiency, the total telecom power inside the waveguide is $0.068\times P_{\mathrm{RefPD}}$ ($0.059\times P_{\mathrm{RefPD}}$) for SHG01 (SHG02). 

For SHG experiments, the SHG power is measured at the visible photodiode (VisPD). In these measurements there is no pinhole in the collection path. Transmisson measurements through the bulk optics for both H and V polarized light (corresponding to the two grating orientations) were measured. Including the grating efficiency, for SHG01 we find the SHG power inside the waveguide corresponds to $114\times P_{\mathrm{VisPD}}$ ($71\times P_{\mathrm{VisPD}}$) for H (V) polarized light, while for SHG2 the power inside the waveguide is $48\times P_{\mathrm{VisPD}}$ ($30\times P_{\mathrm{VisPD}}$) for H (V) polarized light.

\section{Coupling region simulations}
\label{appendix:coupling}

\begin{figure}[h]
\centering\includegraphics[width = \textwidth]{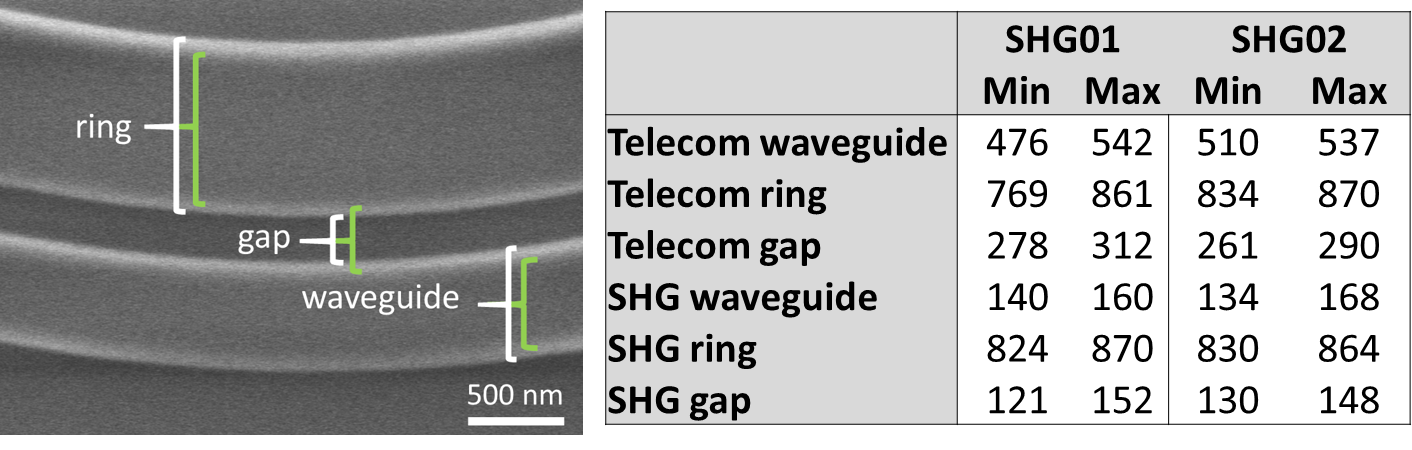} 
\caption{(Left) Device dimensions in the coupling region, shown for the telecom coupling region of device SHG01. White (green) brackets indicate the bottom (top) of the feature.(Right) The measured average dimensions (nm) for each feature for each coupling region for the two devices.} \label{fig:dimensions}
\end{figure}

\begin{figure}[h]
\centering\includegraphics[width = \textwidth]{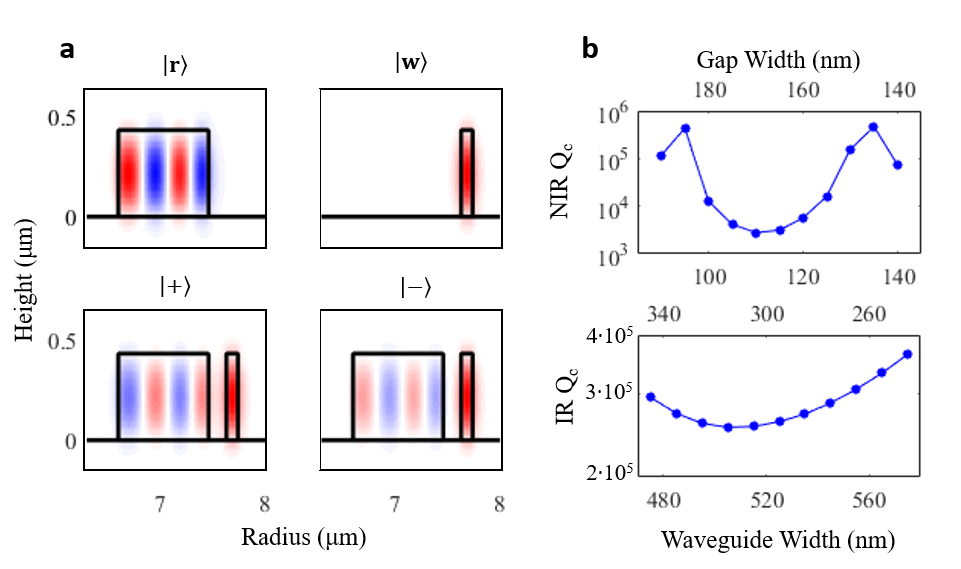} 
\caption{(a) Separate mode profile cross-sections ($\mathrm{\lambda}$ = 775~nm) for the ring resonator ($\ket{r}$) and waveguide ($\ket{w}$) that compose the coupling region. When the two structures are combined, these modes split into two supermodes $\ket{+}$ and $\ket{-}$. (b) Coupling quality factors (logarithmic scale) for IR and NIR modes of a 860~nm wide ring, with gap width (top axis) decreasing as waveguide width (bottom axis) increases. Within measurement uncertainty, wider ring resonators reach slightly lower minimum coupling Q with narrower gaps.}
\label{fig:coupling}
\end{figure}

Simulations of these devices are based on as-fabricated device dimensions measured using scanning electron microscopy (SEM). Measured features are illustrated in Fig. \ref{fig:coupling}. We measured both the top and bottom dimensions for each feature. Fig. \ref{fig:coupling} shows the top dimensions (green brackets) and bottom dimensions (white brackets). Averages of the two measurements are given in Fig. \ref{fig:dimensions}. Assuming a trapezoidal line profile, the measured sidewall angle is $\sim~85$ degrees. Due to this angled sidewall, the top dimensions of the ring and waveguide are smaller than the bottom dimensions and the top dimension of the gap is larger than its bottom dimension. For our device simulations, we used the mean width. 

The coupling efficiency of each wrapped waveguide region was simulated by supermode analysis. As shown in Fig. \ref{fig:coupling}a, if the azimuthal mode number of the curved waveguide mode $\ket{w}$ is similar to the ring resonator mode $\ket{r}$, the two modes combine to form in-phase ($\ket{+}$) and out-of-phase ($\ket{-}$) supermodes when the structures are brought together in the coupling region. Light from the two original modes couples into the supermodes when the waveguide approaches the ring and then couples back into the original modes when the waveguide diverges. The relative phase of the supermodes determines how light is distributed between the original modes. Because the supermodes propagate with distinct azimuthal mode numbers $m^{+}$ and $m^{-}$, the relative phase changes along the coupling region, allowing energy to be transferred from ring to waveguide or vice versa. The field coupling strength $\kappa$ of a single pass through the coupling region is
\begin{equation}
\kappa = \braket{r|+}e^{im^{+}\theta}\braket{+|w} + \braket{r|-}e^{im^{-}\theta}\braket{-|w},
\end{equation}
where $\theta$ is the angular length of the waveguide wrap. In our devices, the telecom (SHG) coupling waveguide wraps around $\mathrm{51^{\circ}}$ ($\mathrm{44^{\circ}}$) of the ring. The resulting coupling quality factor of the ring is
\begin{equation}
Q_c = \frac{4\pi^{2}Rn_g}{\lambda_0 \kappa^2},
\end{equation}
where $R$ is the ring radius, $\lambda_0$ is the free-space wavelength of the resonance, and $n_g$ is the group index of the ring resonator mode. Coupling strengths and quality factors were simulated over the range of measured device dimensions summarized in Fig. \ref{fig:dimensions}. As shown in Fig. \ref{fig:coupling}b, coupling to the second-harmonic mode is very sensitive to variations in device dimensions. Within the range of device measurements, $Q_c$ can vary from $1.2 \times 10^3$ to $1 \times 10^6$, indicating that this mode may be either overcoupled or undercoupled in each device. The telecom coupling region is less sensitive. Most configurations within the range of measurement variations yielded $Q_c$ between $1.5 \times 10^5$ and $5 \times 10^5$. Since this entire range is significantly larger than the measured loaded quality factors, the telecom mode is likely undercoupled in both devices. Finite-difference time-domain simulations of selected devices gave similar results for the plausible range of quality factors.

\section*{Acknowledgements}
This material is based on work supported by the National Science Foundation under award no. 1640986 and  DMR-1454836. We thank Taylor Fryett with assistance in optical testing, and N. Shane Patrick for advice on electron beam lithography. Part of this work was conducted at the Washington Nanofabrication Facility, a National Nanotechnology Coordinated Infrastructure (NNCI) site at the University of Washington, which is supported in part by funds from the National Science Foundation (awards NNCI-1542101, 1337840 and 0335765), the National Institutes of Health, the Molecular Engineering \& Sciences Institute, the Clean Energy Institute, the Washington Research Foundation, the M. J. Murdock Charitable Trust, Altatech, ClassOne Technology, GCE Market, Google and SPTS.

\bibliography{Master}

\end{document}